\documentclass[pra,aps,twocolumn,showpacs,10pt,floatfix]{revtex4-1}
\usepackage{amsmath,amssymb,graphicx,bm,color,ulem}
\usepackage{amsfonts,amsthm}

\begin{document}

\title{Transverse orbital angular momentum of spatiotemporal optical vortices}
\author{Miguel A. Porras}

\affiliation{Grupo de Sistemas Complejos, ETSIME, Universidad Polit\'ecnica de Madrid, R\'{\i}os Rosas 21, 28003 Madrid, Spain}

\begin{abstract}
Spatiotemporal optical vortices (STOVs) are electromagnetic wave packets that transport a phase line singularity perpendicular to their propagation direction. We address the problem of the transverse orbital angular momentum (OAM) ``per photon" actually transported by STOVs  propagating in free space or non-dispersive media, the most frequent experimental situation. Unlike longitudinal vortices in monochromatic light beams, STOVs do not carry any net transverse OAM about a fixed transverse axis crossing its center. However, STOVs transport an intrinsic transverse OAM per photon about a moving, transverse axis through its center, and an opposite extrinsic transverse OAM. Their applications would thus preclude setting particles at rest into rotation, but STOVs could transmit their intrinsic transverse OAM to photons of other waves. The intrinsic transverse OAM per photon of an elliptically symmetric STOV of frequency $\omega_0$ and topological charge $l$ is $\gamma l/2\omega_0$, where $\gamma$ is the STOV ellipticity. Thus circularly symmetric STOVs ($\gamma=1$) carry half the intrinsic longitudinal OAM of circularly symmetric monochromatic light beams with a vortex of the same $l$ and $\omega_0$. We show that the formula $(\gamma+1/\gamma)l/2\omega_0$ for the intrinsic transverse OAM in Phys. Rev. A 107, L031501 (2023) yields infinite values and is not conserved on propagation for a particular STOV. When STOVs lose their elliptical symmetry upon propagation, they preserve the intrinsic transverse OAM $\gamma l/2\omega_0$ despite the phase singularity may split, the split singularities may disappear, or even change the sign of their topological charges.
\end{abstract}

\maketitle


\setlength {\abovedisplayskip} {6pt plus 3.0pt minus 4.0pt}
\setlength {\belowdisplayskip} {6pt plus 3.0pt minus 4.0pt}

\

\section{Introduction}\label{INTRO}

Within the dynamic area of research on the so-called structured light, optical vortices play a prominent role. Vortices in monochromatic light beams have been studied for decades \cite{SHEN}. They feature a phase line singularity along the beam propagation axis, e.g., the $z$ axis, where the intensity vanishes, surrounded, in their simplest version, by a circularly symmetric distribution of intensity. The orbital angular momentum (OAM) carried by these vortex beams is well-understood, and is commonly quantified by the OAM ``per photon" as $l/\omega_0$, where the integer $l$ is the topological charge of the vortex, and $\omega_0$ the beam frequency. For many applications these vortices are not nested in monochromatic beams, but in ultrafast pulsed beams, which are often called ``spatiotemporal vortices".

However, the above name, or more precisely, spatiotemporal optical vortices (STOVs), is reserved since a few years for pulsed beams carrying a vortex whose phase line singularity is not longitudinal but transverse to the direction of propagation, say the $y$ axis, with the gradient of the phase circulating in the $z$-$x$ plane, or equivalently, in the $t$-$x$ plane in the usual description of pulsed beams at transversal planes $z=\mbox{const.}$ as time goes on.

STOVs were first observed experimentally in optical collapse and filamentation \cite{JHAJJ}, but their linear nature made it possible to generate them in air using standard pulse and beam shaping techniques \cite{HANCOCK,CHONG}. Recent generalizations of these purely transversal STOVs include STOVs with arbitrarily oriented phase line singularity \cite{WANG}, STOVs with mixed phase and polarization singularities \cite{CHEN}, non-diffracting Bessel-type STOVs \cite{CAO}, etc.

Detailed descriptions of the propagation features of STOVs with transverse OAM can be found in \cite{HANCOCK3,HUANG1,HUANG2}, including closed-form expressions for higher-order STOVs ($l>1$) propagating in free space \cite{PORRAS4}. Circularly symmetric, or given the different nature of the $x$ and $t$ coordinates, elliptically symmetric STOVs, are theoretically considered as prototype STOVs, and they are assumed to carry transverse OAM, but different authors provide expressions attributing different amounts of total, intrinsic, and extrinsic transverse OAM per photon \cite{HANCOCK3,BLIOKH1,BLIOKH2}, which has sparked a subtle debate \cite{ROADMAP}. This point is relevant not only theoretically but also to experiments with STOVs where the transverse OAM is intended to be transferred to a second harmonic beam \cite{HANCOCK2,MURNANE}, or even to high harmonics \cite{FANG}, and may be also relevant to other linear or nonlinear interactions with matter \cite{MAZANOV,PORRAS5}.

Barnett \cite{BARNETT} has identified the correct physical magnitude in classical, Maxwellian electromagnetism that can be identified with the angular momentum transported by a monochromatic vortex beam: the angular-momentum flux crossing a transversal plane, or angular momentum per unit time, since the total angular momentum is infinite for a continuous beam. This formulation allows a physically meaningful separation of spin and orbital angular momentum even for nonparaxial beams \cite{BARNETT}. Here, we adopt this formulation and extend it to waves localized in space and also in time, such as STOVs, and identify the angular momentum carried by the STOV with the angular momentum flux integrated in time. Considering linearly polarized wave packets, this angular momentum is further identified with the OAM carried by the STOV. We find that the OAM with respect to any static transverse axis traversing the center of the STOV is zero. Yet, the STOV carries an OAM with respect to a moving axis permanently traversing the center of the STOV, which we identify as an intrinsic OAM, and an opposite extrinsic OAM with respect to the static axis.

At the transversal plane where a STOV is elliptic, the intrinsic OAM per photon of a STOV can be calculated as $l\gamma/2\omega_0$, where $l$ is the topological charge of the STOV at that plane, and $\gamma$ measures the ellipticity. This yields $l/2\omega_0$ for a circular STOV, half the OAM of circular spatial vortices. An intuitive explanation of this fact is provided. When the STOV loses its elliptical symmetry due to diffraction effects, the intrinsic OAM is conserved, but there is no any relationship between the OAM and the topological charge(s) of the vortices in the STOV, which may disappear, even reverse their sign.

\section{transverse OAM transported by electromagnetic wave packets}\label{TRANS}

In classical electromagnetic theory, the energy $W$, momentum $\vec P$ and angular momentum $\vec J$ carried by an electromagnetic wave can be determined from their conservation laws. Conservation of energy is expressed by the continuity equation for energy $\partial_t w + \partial_m S_m=0$, where $w=(1/2)(\varepsilon_0|{\vec E}|^2 +\mu_0^{-1}|{\vec B}|^2)$ is the energy density $\vec E, \vec B$ the real-valued electric and magnetic vectors, $\varepsilon_0$ and $\mu_0$ the electric permittivity and magnetic permeability of vacuum, and ${\vec S} =\mu_0^{-1}{\vec E}\times {\vec B}$ is the Poynting vector, or energy flux density. The repeated subindex $m$ implies summation over all its values, $m=x,y,z$, the divergence operator in this case. Analogously, conservation of each component of momentum reads $\partial_t p_i + \partial_m T_{im}=0$, where $p_i=S_i/c^2$ is the $i$ component of the momentum density, and
$T_{im}=(1/2) \delta_{im}(\varepsilon_0|\vec E|^2 + \mu_0^{-1}|\vec B|^2) - \varepsilon_0 E_i E_m - \mu_0^{-1}B_iB_m $
is the momentum flux density, with $\delta_{im}$ the Kronecker delta. The $im$ component is the flux of the $i$ component of the momentum across an infinitesimal surface perpendicular to the $m$ direction. These continuity equations are direct consequences of Maxwell equations \cite{BARNETT}.

There is a third continuity equation consequence of the conservation of angular momentum. We express the Cartesian components of the angular momentum density $\vec j=\vec r \times \vec p$ more efficiently as $j_i=\epsilon_{ijk}x_jp_k$, where $\epsilon_{ijk}$ is the permutation symbol of values $\epsilon_{ijk}=+1$ if $ijk = 123, 312, 231$,  $\epsilon_{ijk}=-1$ if $ijk = 321, 132, 213$, and zero otherwise. As shown in \cite{BARNETT}, the continuity equation for angular momentum is
\begin{equation}
\partial_t j_i +\partial_{m} M_{im}=0\,,
\end{equation}
where
\begin{equation}
M_{im}=\epsilon_{ijk}x_jT_{km}
\end{equation}
is the angular momentum flux density, yielding the flux density of the $i$ component of the angular momentum across a surface perpendicular to the $m$ direction, and having units of angular momentum per unit area and unit time.

As for energy and momentum, the continuity equation can be expressed via divergence's theorem in integral form as
\begin{equation}\label{CONSER}
\frac{d}{dt}\int_V j_i dV = -\oint_S M_{im} dS_m\,,
\end{equation}
meaning that the variation of the angular momentum in a volume $V$ equals to the inward angular momentum flux across its surface $S$, i.e., the angular momentum entering into $V$ from outside. For a transversally and temporally localized wave packet propagating along the $z$ axis, we take $V$ as a coaxial cylinder of bases at planes $z_1$ and $z_2>z_1$. If the radius of the cylinder tends to infinity, the flux across its lateral surface vanishes, and (\ref{CONSER}) reads
\begin{equation}
\frac{d}{dt}\int_V j_i dV =\int_{z_1} M_{iz} d\vec x_\perp- \int_{z_2} M_{iz} d\vec x_\perp\,,
\end{equation}
since $dS_z=dxdy\equiv d\vec x_\perp$, with the integral in $x,y$ covering the entire transversal planes. Integration also in time from $t=-\infty$ to $t=+\infty$ yields
\begin{equation}
\int_{z_1} M_{iz} d\vec x_\perp dt - \int_{z_2} M_{iz} d\vec x_\perp dt =\left.\int_V j_i dV\right]_{t=-\infty}^{t=\infty} \,.
\end{equation}
The right hand side vanishes since at $t=-\infty$ and at $t=+\infty $ there is no angular momentum in the limited volume $V$ from $z_1$ to $z_2$ for a temporally localized wave. The total angular momentum that crosses any transversal section as the wave packet surpasses that section,
$J_i=\int M_{iz} d\vec x_\perp dt $, is then independent of $z$. In \cite{BARNETT}, Barnett identified the angular momentum per unit time carried by a monochromatic light beam with the angular momentum flux ${\cal M}_i=\int M_{iz} d\vec x_\perp$ through a transversal section $z$. Accordingly, we identify here the total angular momentum carried by a beam localized in time with the angular momentum flux ${\cal M}_i$, or angular momentum per unit time, integrated to all times, i.e., $J_i=\int M_{iz} d\vec x_\perp dt $.

We wish to apply these fundamental relations to wave packets such as STOVs supposedly carrying transverse OAM along a transversal direction, say, the $y$ direction. The angular momentum flux density and angular momentum about the $y$ axis, i.e., about the axis $(x,z)=0$, are
\begin{equation}\label{MYZ}
M_{yz} =  z T_{xz} -x T_{zz}
\end{equation}
and
\begin{equation}\label{JY}
J_y= \int M_{yz} d\vec x_\perp dt,
\end{equation}
where
\begin{eqnarray}
T_{xz}&=&-\varepsilon_0 E_xE_z - \mu_0^{-1} B_x B_z\label{TXZ}, \\
T_{zz}&=& \frac{1}{2}\left[\varepsilon_0(E_x^2 \!+\! E_y^2 \!-\! E_z^2)+\mu_0^{-1}(B_x^2 \!+\! B_y^2 \!-\! B_z^2)\right] \label{TZZ}.
\end{eqnarray}
Since we will only consider linear polarization, this angular momentum will be identified as the transverse OAM carried by the wave packet.

To the purpose of computing the transverse OAM ``per photon," we also consider the energy transported by the wave packet along the $z$ direction. A procedure identical (and much better known) as above, but starting with the continuity equation for energy, yields the carried energy as
\begin{equation}
W= \int S_z d\vec x_\perp dt \,,
\end{equation}
where $S_z=\mu_0^{-1}(E_xB_y -E_yB_x)$ is the energy flux density across the transversal plane. Obviously, $W$ is also independent of $z$.

\section{Electromagnetic fields of paraxial and quasimonochromatic wave packets} \label{ELECTRO}

In current experiments STOVs propagate paraxially and their duration is much longer than the carrier period $2\pi/\omega_0$, i. e., they are many-cycle, quasimonochromatic, or narrowband wave packets, of typical duration in the scale of hundred of femtoseconds in the visible spectrum. Under these conditions, the propagation of the complex envelope $\psi(x,y,z,t')$ of any transversal component of the complex electromagnetic field, $\psi(x,y,z,t') e^{-i\omega_0 t'}$, where $t'=t-z/c$ is the local time, and $\omega_0$ is the carrier frequency, is accurately described by the linear Schr\"odinger equation
\begin{equation}
\partial_z \psi = \frac{i}{2k_0}\Delta_\perp \psi\,,
\end{equation}
where $\Delta_\perp=\partial^2_x+\partial^2_y$, and $k_0=\omega_0/c$ is the propagation constant. This is the common approximation to paraxial propagation of narrowband pulses, ruled by $\partial_z \psi = (i/2k_0)\Delta_\perp \psi -i(k_0^{\prime\prime}/2)\partial^2_{t'}\psi$, when group velocity dispersion $k_0^{\prime\prime}$ vanishes. A more detailed derivation of the Schr\"odinger equation for paraxial and quasimonochromatic fields can be found in \cite{PORRAS4,HEYMAN,BESIERIS} and references therein.

Following Lax's {\it et al} perturbation theory \cite{LAX}, the paraxial electromagnetic fields can be constructed from solutions of the Schr\"odinger equation as
\begin{eqnarray}\label{FIELDS}
E_x&=&\mbox{Re}\left\{\psi e^{-i\omega_0 t'}\right\}, \,\,\,\, E_z  =\mbox{Re}\left\{\frac{i}{k_0}\partial_x\psi e^{-i\omega_0 t'}\right\},  \\
B_y&=& \mbox{Re}\left\{\frac{1}{c}\psi e^{-i\omega_0t'}\right\}, B_z  = \mbox{Re}\left\{\frac{i}{k_0 c}\partial_y\psi e^{-i\omega_0 t'}\right\}, \nonumber
\end{eqnarray}
$E_y=0$, and $B_x=0$ for linear polarization along $x$. For linear polarization along $y$, exchange $x\leftrightarrow y$ in all of the above equations.

\section{Transverse orbital angular momentum of paraxial and quasimonochromatic wave packets.} \label{OAM}

For quasimonochromatic light, integrations in time from $-\infty$ to $+\infty$ to obtain the transverse OAM and the energy can be evaluated in two steps. First, when the fields (\ref{FIELDS}) are introduced in (\ref{TXZ}) and (\ref{TZZ}) terms oscillating at $2\omega_0$ are cancelled upon integration over a carrier period, and only those independent of $\omega_0$ remain, leading to the cycle-averaged transverse OAM flux density and transverse OAM as
\begin{equation}\label{MYZCYCLE}
\langle M_{yz}\rangle =z \langle T_{xz}\rangle - x \langle T_{zz}\rangle, \quad J_y= \int \langle M_{yz}\rangle d\vec x_\perp dt\,,
\end{equation}
where
\begin{eqnarray}
\langle T_{xz} \rangle &=& \frac{\varepsilon_0}{2k_0}\mbox{Im}\{\psi^\star \partial_x \psi\} = \frac{\varepsilon_0}{2k_0} A^2 \partial_x\Phi ,\label{TXZCYCLE} \\
\langle T_{zz} \rangle &=& \frac{1}{2}\varepsilon_0 |\psi|^2  =\frac{1}{2}\varepsilon_0 A^2 , \label{TZZCYCLE}
\end{eqnarray}
regardless polarization is along $x$ or along $y$, and where we have introduced the amplitude $A$ and phase $\Phi$ of the complex envelope $\psi=Ae^{i\Phi}$ in the second set of equations.

Also, using the fields in (\ref{FIELDS}) in $S_z=\mu_0^{-1}(E_xB_y -E_yB_x)$, the cycle-averaged $z$-component of the energy flux density is
$\langle S_z \rangle= (1/2) \varepsilon_0 c |\psi|^2 = (1/2) \varepsilon_0 c A^2$. The energy transported by the wave packet is then
\begin{equation}\label{WCYCLE}
W=\frac{1}{2}\varepsilon_0 c \int A^2 d\vec x_\perp dt'\,.
\end{equation}

We may decompose the transverse OAM flux density and the transverse OAM into intrinsic and extrinsic contributions. The intrinsic part is associated with the transverse OAM about a moving axis parallel to the $y$ axis traversing the wave packet ``center", and the extrinsic part is associated with the rotation of this center about the $y$ axis, i.e., about $(x,z)=0$. For a wave packet moving at $c$, as STOVs, the moving axis at the plane $z$ is $[x-x_m,z-c(t-t'_m)]=0$, or in terms of the local time $[x-x_m,-c(t'-t'_m)]=0$, where $x_m$ and $t'_m$ define the wave packet center at the plane $z$, and take into account that at the plane $z$, the $x$-center $x_m$ may not be $x=0$ and that the temporal center may be delayed from $t=z/c$ by $t'_m$ for a complex wave packet lacking symmetries. The intrinsic and extrinsic transverse OAM flux densities are then
\begin{eqnarray}
\langle M_{yz}^{(i)}\rangle&=& [z-c(t-t'_m)]\langle T_{xz}\rangle -(x-x_m)\langle T_{zz}\rangle \nonumber \\
                           &=& -c(t'-t'_m) \langle T_{xz}\rangle  - (x-x_m)\langle T_{zz}\rangle \,,\label{MYZINT} \\
\langle M_{yz}^{(e)}\rangle &=& c(t-t'_m) \langle T_{xz}\rangle  - x_m \langle T_{zz} \rangle \nonumber \\
                           &=& [z+ c(t'-t'_m)] \langle T_{xz}\rangle  - x_m \langle T_{zz} \rangle\,, \label{MYZEX}
\end{eqnarray}
verifying $\langle M_{yz}^{(i)}\rangle+ \langle M_{yz}^{(e)}\rangle=\langle M_{yz}\rangle $, with the intrinsic and extrinsic transverse OAM given by
\begin{equation}\label{JYINTEX}
J_y^{(i)}=\int \langle M_{yz}^{(i)}\rangle d\vec x_\perp dt',\quad J_y^{(e)}=\int \langle M_{yz}^{(e)}\rangle d\vec x_\perp dt' \,,
\end{equation}
also verifying $J_y^{(i)}+J_y^{(e)}=J_y$. Note that integration to all times $t$ yields the same result as integration in $t'$. Let us remark that the parameters $x_m$ and $t'_m$ are not directly related to the structure of the wave packet as seen in space $(x,y,z)$ at different times, but just describe what happens at a particular plane $z$. Accordingly, the natural definitions of $x_m$ and $t'_m$ that are consistent with our formulation are $x_m= W^{-1}\int \langle S_z\rangle x d\vec x_\perp dt'$ and $t'_m = W^{-1}\int \langle S_z\rangle t' d\vec x_\perp dt'$, or equivalently,
\begin{equation}
x_m = \frac{\int A^2 x d\vec x_\perp dt'}{\int A^2  d\vec x_\perp dt'},\quad
t'_m =\frac{\int A^2 t' d\vec x_\perp dt'}{\int A^2  d\vec x_\perp dt'}.
\end{equation}
(One could think on relating these parameters to the energy density $\langle w\rangle$, but the result would be the same, since $\langle w \rangle$ and $\langle S_z\rangle$ are proportional for paraxial fields.)

Using all the above expressions, the total, intrinsic, and extrinsic transverse OAM carried by the wave packet are given by
\begin{eqnarray}\label{JYCYCLE}
J_y&=&\frac{\varepsilon_0 z}{2k_0}\!\!\int \!\!\!A^2\partial_x\Phi d\vec x_\perp dt' -\frac{1}{2}\varepsilon_0 \!\!\int\!\!\! A^2 x d\vec x_\perp dt' \,, \\
J_y^{(i)}&=&-\frac{\varepsilon_0 c}{2k_0}\!\!\int\!\!\! A^2\partial_x\Phi (t'-t'_m)d\vec x_\perp dt' \label{JYINTCYCLE} \,, \\
J_y^{(e)}&=&\frac{\varepsilon_0 z}{2k_0}\!\!\int \!\!\!A^2\partial_x\Phi d\vec x_\perp dt' + \frac{\varepsilon_0 c}{2k_0}\!\!\int\!\!\! A^2\partial_x\Phi (t'\!-\!t'_m)d\vec x_\perp dt' \nonumber \\
         &-&\frac{1}{2}\varepsilon_0 \!\!\int\!\!\! A^2 x d\vec x_\perp dt' \,.\label{JYEXCYCLE}
\end{eqnarray}
It should be clear that $J_y$ and $J_y^{(e)}$ are referred to the $(x,z)=0$ axis, and would take other values if another transverse axis is taken. However, $J_y^{(i)}$ would not change since it is always referred to the wave packet center. Indeed, the transverse OAM about a new transverse axis $(x,z)=(x_0,0)$ is the same as the transverse OAM about $(x,y)=0$ of the wave packet translated by $-x_0$. Replacing $A(x,y)$ and $\Phi(x,y)$ with $A(x+x_0,y)$ and $\Phi(x+x_0,y)$ in (\ref{JYCYCLE}), (\ref{JYINTCYCLE}) and (\ref{JYEXCYCLE}), changing to variables $x'=x+x_0$ and $y'=y$, and using that $\partial_{x'}=\partial_x$, one immediately obtains the new OAMs as $J_y(x_0)=J_y +x_0 P_z$, $J_y^{(i)}(x_0)= J_y^{(i)}$, and $J_y^{(e)}(x_0)=J_y +x_0 P_z$, where $P_z =\int \langle T_{zz}\rangle d\vec x'_\perp dt'$ is the $z$ component of the momentum carried by the wave packet.

\section{Transverse orbital angular momentum of spatiotemporal vortices} \label{VORTICES}

As is well-known, monochromatic light beams with circular symmetry except for an azimuthal phase dependence $e^{il\phi}$, $\phi=\tan^{-1}(y/x)$  carry a longitudinal orbital angular momentum (OAM) per photon $J_z/W = J_z^{(i)}/W= l/\omega_0$ about the $z$-axis $(x,y)=0$, an OAM that is purely intrinsic.

The amount of transverse OAM carried by STOVs is a subject of debate. Gaussian-type STOVs considered theoretically have a transversal plane $z$ where they present elliptical symmetry in the $t'$-$x$ plane \cite{HANCOCK3}. Bessel-type STOVs are elliptical everywhere \cite{BLIOKH1,BLIOKH2}. We take advantage of the conservation of transverse OAM (as verified later) to evaluate the transverse OAM at that plane.

With elliptical symmetry, STOVs are of the form
\begin{equation}
\psi = f(\rho) e^{-i l\varphi}\,,
\end{equation}
where $\rho = \sqrt{\tau^2+\xi^2}$ and $\varphi = \tan^{-1}(\xi/\tau)$, with $\tau=t'/t_0$, $\xi=x/x_0$, are polar coordinates in the spatiotemporal plane $t'/t_0$-$x/x_0$, and the parameters $t_0$ and $x_0$ determine the ellipticity $\gamma=ct_0/x_0$. The function $f(\rho)$ may be complex and behaves as $\rho^{|l|}$ close to $\rho=0$. Observing that $t'=t-z/c$, $e^{-i l \varphi}$ with positive $l$ corresponds to a vortex in which the phase increases counterclockwise as viewed in the $z$-$x$ plane, as for a spatial vortex in the $x$-$y$ plane with positive topological charge. Also, the STOV $\psi$ should be accompanied by an arbitrary complex amplitude profile $Y(y)$ along the $y$ direction, but this factor is omitted in $\psi$ since it only yields $|Y(y)|^2$ factors in all the densities, and factorized integrals of $|Y(y)|^2$ on $y$ in the transverse OAM and energy that cancel when evaluating the transverse OAM per photon.

Using that $A=|f(\rho)|$, $\Phi=\mbox{arg}[f(\rho)] - l\varphi$ and $t_m=0$, and changing to polar coordinates in the integrals in  (\ref{JYCYCLE}, \ref{JYINTCYCLE}, \ref{JYEXCYCLE}), and (\ref{WCYCLE}) (without the integrals in $y$), it is a straightforward calculation to arrive at
\begin{eqnarray}
J_y\,\,\,&=&0, \\
J_y^{(i)} &=& l \frac{\varepsilon_0 c t_0^2}{2k_0} \pi \int_0^{\infty} |f(\rho)|^2 \rho d\rho , \\
J_y^{(e)} &=& -l \frac{\varepsilon_0 c t_0^2}{2k_0} \pi \int_0^{\infty} |f(\rho)|^2 \rho d\rho ,
\end{eqnarray}
and
\begin{equation}
W= \varepsilon_0 c x_0t_0 \pi \int_0^\infty |f(\rho)|^2\rho  d\rho\,.
\end{equation}
Thus, STOVs do not carry any net transverse OAM with respect to the $y$-axis $(x,z)=0$, but have opposite intrinsic and extrinsic transverse OAM. The total, intrinsic and extrinsic transverse OAM per photon are obtained to be
\begin{equation}\label{JYW}
\frac{J_y}{W}=0\,, \quad \frac{J_y^{(i)}}{W} = \frac{l}{2}\frac{\gamma}{\omega_0}\,, \quad \frac{J_y^{(e)}}{W} = - \frac{l}{2}\frac{\gamma}{\omega_0}\,.
\end{equation}
Of course the first and last relation only hold for a fixed transversal axis passing through the STOV center, but the expression for the intrinsic OAM is independent of the choice of the axis, as demonstrated above.
The second equation coincides with the intrinsic transverse OAM in \cite{HANCOCK3} for Gaussian-type STOVs in vacuum, and apply also here to Bessel-type STOVs. A different expression replacing $\gamma$ with $\gamma + 1/\gamma$ has been proposed for the intrinsic transverse OAM in \cite{BLIOKH1,BLIOKH2}, where the second equation in (\ref{JYW}) is identified as the total OAM.

\begin{figure}
  \centering
  \includegraphics[width=8cm]{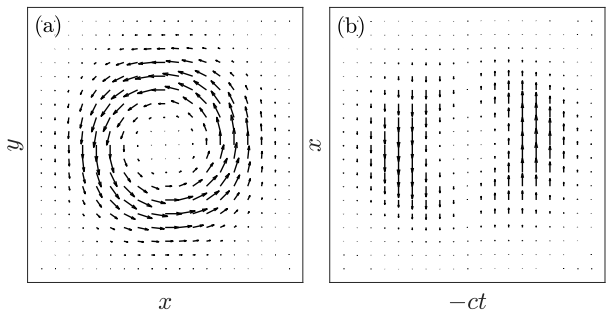}
  \caption{Schematic drawing of the cycle-averaged components of the momentum density, $\langle p_x\rangle$ and $\langle p_y\rangle$, for (a) a spatial vortex and (b) a STOV, both with topological charge $l=3$. The horizontal axis is $-ct$ to visualize the rotation in $z$-$x$ space.}\label{Fig1}
\end{figure}

When the STOV is round ($\gamma=1$), the intrinsic transverse OAM per photon is $l/2\omega_0$, half of the longitudinal OAM of spatial vortices. The formula with $\gamma + 1/\gamma$ in \cite{BLIOKH1,BLIOKH2,ROADMAP} yields $l/\omega_0$ for round STOVs, which looks more appealing since it coincides with the longitudinal OAM. However, in addition to be supported by classical electromagnetic theory, the fact that circular STOVs carry half of the longitudinal OAM is even intuitively understandable from Fig. \ref{Fig1}. In these symmetric STOVs, the $z$ component of the momentum flux density $\langle T_{zz}\rangle = c \langle p_z\rangle$ does not contribute to the transverse OAM, and the same happens for spatial vortices. Figure \ref{Fig1} depicts only the components that contribute in each case. In standard vortices there are two linear momentum fluxes, $\langle T_{xz}\rangle =c\langle p_x\rangle$ and $\langle T_{yz}\rangle= c\langle p_y\rangle $, contributing equally to the longitudinal OAM, as in (a), while in STOVs only $\langle T_{xz}\rangle =c \langle p_x\rangle$ contributes to the transverse OAM, as in (b), making understandable the factor $1/2$. Similar arguments were presented in \cite{HANCOCK3} to support the factor $1/2$.

\begin{figure}
  \centering
  \includegraphics[width=5.5cm]{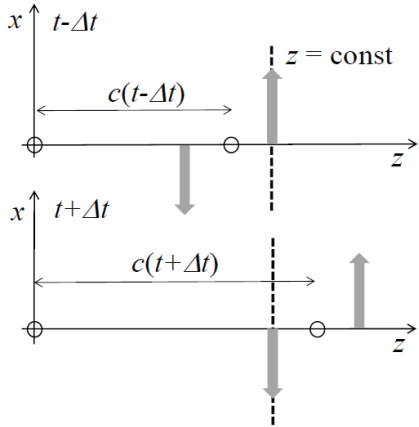}
  \caption{Understanding the total, intrinsic and extrinsic transverse OAM of STOVs.}\label{Fig2}
\end{figure}

The sketch in Fig. \ref{Fig2} is aimed at visualize more intuitively the above results on the transverse OAM. The arrows symbolize the positive and negative linear momenta along $x$ transported by the STOV in its leading and rear parts when the STOV is positively charged, as in Fig. \ref{Fig1}(b). By the same symmetry reasons as above the linear momentum along $z$ does not contribute to the transverse OAM. The STOV is shown at the instant of time $t-\Delta t$ at which the positive momentum $P_x$ flows through the plane $z$, and at the instant of time $t+\Delta t$ at which the negative momentum $-P_x$ flows. The $y$ axis $(x,z)=0$ and the moving $y$ axis $(z-ct,x)=0$ are indicated by small circles.
At $t-\Delta t$ the leading part contributes with a positive transverse OAM $zP_x$, and at $t+\Delta t$ the rear part contributes with the opposite transverse OAM $-zP_x$, by which the STOV does not carry transverse OAM with respect to the $y$ axis $(x,z)=0$. In contrast, at $t-\Delta t$ the transverse OAM contribution with respect to the STOV center is $c\Delta t P_x$, and the transverse OAM contribution at $t+\Delta t$ is again $c\Delta t P_x$, making $2c\Delta t P_x$ in total. The extrinsic transverse OAM refers to the transverse OAM of the STOV center with respect to the $y$ axis $(x,z)=0$. At $t-\Delta t$ this center is at $c(t-\Delta t)$, so that the contribution to the transverse OAM is $c(t-\Delta t)P_x$. At $t+\Delta t$, the center is at $c(t+\Delta t)$, yielding a transverse OAM $-c(t+\Delta t)P_x$. The sum of these two is $-2c\Delta t P_x$, just the opposite of the intrinsic transverse OAM.

Elliptical STOVs experience drastic changes on propagation in a non-dispersive medium, including the loss of elliptical symmetry and even the disappearance of the vortices. Even if the total transverse OAM is zero and conserved, one may suspect that the intrinsic transverse OAM may change on propagation, along with the opposite change in the extrinsic transverse OAM. Also, the total and the intrinsic OAMs are different, and therefore the conservation of the latter cannot be taken for granted. However, the separation of intrinsic and extrinsic transverse OAM in paraxial and quasimonochromatic wave packets propagating according to the Schr\"odinger equation is robust: Starting with $dJ_y/dz$ (by way of verification) and $dJ_y^{(i)}/dz$, with $J_y$ and $J_y^{(i)}$ given by (\ref{JYCYCLE}) and (\ref{JYINTCYCLE}) for general $\psi$, using
$A^2 \partial_x\Phi= \mbox{Im}\{\psi^\star \partial_x \psi\}$, and $A^2 =\psi\psi^\star$ for convenience, introducing the derivative with respect to $z$ into the integrals, using Schr\"odinger equation to evaluate $\partial\psi/\partial z$, and performing several integrations by parts, all integrals are found to vanish, and hence $dJ_y/dz=0$ and $dJ_y^{(i)}/dz=0$. The intrinsic transverse OAM is then also conserved on propagation.

\begin{figure*}[t]
  \centering
  \includegraphics[width=17cm]{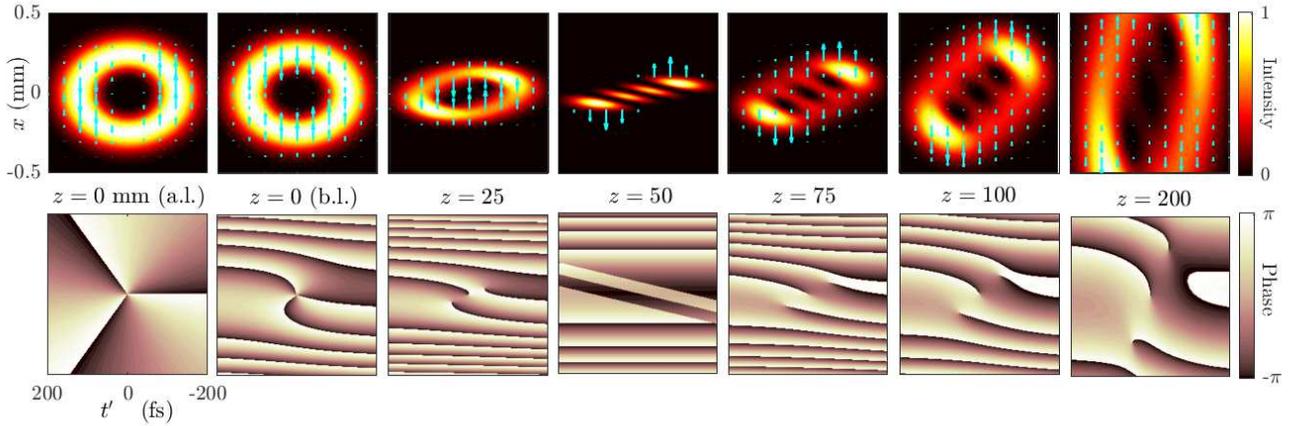}
  \caption{Propagation in vacuum of the focused STOV $\psi=[(t'/t_0) - i (x/x_0)]^{|l|} e^{-(x/x_0)^2 - (t/t_0)^2}e^{-ik_0x^2/2f}$, of positive topological charge $l=3$, with $x_0=0.2$ mm, $t_0=100$ fs, $f=50$ mm, of carrier frequency $\omega_0=2.5$ rad/fs ($\lambda_0=754$ nm), evaluated by solving numerically Schr\"odinger equation. First row: Cycle-averaged intensity $\langle S_z\rangle$ (contour plot) and momentum density $\langle p_x\rangle$ (arrows) in the plane $t'$-$x$ at the indicated propagation distances. Second row: the same for the phase $\mbox{arg}\{\psi\}$. The intensity is normalized to its peak value at each distance. The momentum is relative to its maximum value. a.l.: after the lens, b.l.: before the lens. The order of time is reversed to visualize rotations in $z$-$x$ space.}\label{Fig3}
\end{figure*}

\section{Examples and discussion}\label{EXAMPLES}

The above results would seem to indicate that the intrinsic transverse OAM is directly connected to the topological charge of the STOV, but the relation $J_y^{(i)}/W=l\gamma/2\omega_0$ holds only at the plane (or planes) where the STOV is elliptic. Indeed, the spatiotemporal singularity of the phase is a dark region that itself does not transport energy, momentum or angular momentum. It is only the momentum flux density about the STOV center that determines the intrinsic transverse OAM. Propagation of STOVs may maintain the sign of the topological charge, eliminate the vortices, or reverse their sign, while the intrinsic transverse OAM is conserved. Reversal of the sign has been previously described in STOVs in normally dispersive media \cite{HANCOCK3}, in STOVs in free space \cite{PORRAS4}, and is a phenomenon common to spatial vortices, e.g., in nonlinear media \cite{PORRAS6}, observed in free space as earlier as in \cite{TORNER}.

For example, the elliptic STOV $\psi=[(t'/t_0) - i (x/x_0)]^{|l|} e^{-(x/x_0)^2 - (t/t_0)^2}$ of positive charge $l$ and transverse OAM $J_y^{(i)}/W=l\gamma/2\omega_0$ at $z=0$ continues to have a total positive charge $l$ in $l$ split vortices of unity charge on propagation \cite{PORRAS4}. However, an elliptic STOV of positive charge $l$ converging from $z=-\infty$ to a focus transforms at $z=+\infty$ into an elliptic STOV of negative charge $-l$ \cite{PORRAS4}. The intrinsic transverse OAM of this STOV is zero \cite{PORRAS4}; indeed $J_y^{(i)}/W=l\gamma/2\omega_0=0$ at $z= \pm \infty$ since $\gamma=ct_0/x_0=0$ given the constant value of $t_0$ and indefinitely increasing value of $x_0$ as $z\rightarrow\pm\infty$. It is then clear that spatiotemporal phase singularities in STOVs may not be associated with any transverse OAM.

In passing, we note that the expression $J_y^{(i)}/W=l (\gamma+1/\gamma)/2\omega_0$ in \cite{BLIOKH1,BLIOKH2,ROADMAP} and more recently in \cite{BLIOKH3} for the intrinsic transverse OAM would yield, in the above example, $J_y^{(i)}/W= +\infty$ for the elliptic converging STOV from $z=-\infty$ and $J_y^{(i)}/W=-\infty$ for the diverging elliptic STOV to $z=+\infty$, which is not conserved and hard to justify.

Yet, it could still be argued that flipping of topological charge is allowed because the transverse OAM is zero.
In the example of Fig. \ref{Fig3} the STOV carries positive intrinsic transverse OAM, but still the positive topological charges turns negative. The expression  $\psi=[(t'/t_0) - i (x/x_0)]^{|l|} e^{-(x/x_0)^2 - (t/t_0)^2}e^{-ik_0x^2/2f}$, $f>0$, represents a elliptical STOV with a converging spherical wavefront. According to (\ref{JYINTCYCLE}), the positive transverse OAM $J_y^{(i)}/W=l\gamma /2\omega_0$ is not altered by the introduction of the converging wave front, but only changes the wave front and momentum density distribution (first and second columns in Fig. \ref{Fig3}). It is clear from the second row that the $l$-charged vortex splits in $l$ unit-charged vortices (columns 1, 2 and 3), they disappear at the focus (column 4), and new $l$ unit-charged of opposite sign emerge after the focus (columns 5, 6, and 7). We note that an elliptical symmetric $-l$-charged STOV is not formed at any distance, including the far field (last column). An elliptical STOV at the far field would imply, as in the previous example, that the transverse OAM is zero, but it continues to be $J_y^{(i)}/W=l\gamma/2\omega_0$. This example also illustrates that the intrinsic transverse OAM cannot be directly associated with the ``rotation" of the intensity pattern. Before the focus, this pattern appears to rotate clockwise, but after the focus it does counterclockwise. An intuition of the sign of the intrinsic transverse OAM can only be obtained by visualizing $\langle p_x\rangle$ (blue arrows in the first row).

\section{Conclusion}\label{CONCLU}

In short, STOVs in non-dispersive media do not carry any OAM about a fixed transversal axis crossing its center, but carry opposite intrinsic and extrinsic transverse OAM in half the amount of the longitudinal OAM carried by spatial vortices with the same topological charge, to fix ideas with the simplest case of a circular STOV. When the STOV is circular or elliptic, the intrinsic OAM and the topological charge of the spatiotemporal phase singularity have the same sign and are proportional; otherwise STOVs with positive or negative intrinsic transverse OAM can indistinctly feature spatiotemporal phase singularities with topological charge of equal or opposite sign.

Probably the discrepancies with other authors and among themselves has to do with subtle differences in the respective formalisms. Here we have considered a STOV as a classical wave subject to the laws of classical electromagnetism. Treating angular momentum as a quantum mechanical operator that acts on the wave field \cite{HANCOCK3}, or invoking the concept of ``photon wave function" \cite{BIRULA} may be problematic in very special situations. In this sense we have used the expression ``OAM per photon" as a simple abbreviation of ``OAM carried per unit energy carried," regardless of whether it is true at the quantum level or not.

We have limited our analysis to STOVs in non-dispersive media, as air with the long durations of these STOVs, to focus on addressing the controversy, because STOVs are mostly generated and propagate in this medium, and because the present results can impact the results and interpretation of experiments where STOVs interact with matter. Our analysis do not apply to STOVs in dispersive media, but can be understood as a confirmation of, and used to reinterpret, the results of the analysis of the OAM content in non-dispersive and dispersive media in \cite{HANCOCK3}. We mention that STOVs were conceived theoretically in nonlinear media as early as in \cite{BORIS}, and that the present analysis may also shed light on the OAM content of those nonlinear STOVs.

\section*{ACKNOWLEDGMENT}

This work has been partially supported by the Spanish Ministry of Science and Innovation, Gobierno de España, under Contract No. PID2021-122711NB-C21. The author also acknowledges support as visiting professor of La Sapienza University and Dipartimento di Fisica of La Sapienza.

\end{document}